\documentclass[conference,10pt]{IEEEtran}
\IEEEoverridecommandlockouts
\pdfoutput=1
\usepackage{graphicx, epstopdf, float}
\usepackage{amssymb}
\usepackage{algorithm}
\usepackage{algorithmic}
\usepackage[cmex10]{amsmath}
\usepackage{amsthm}
\usepackage{mathrsfs}
\usepackage{cite}
\usepackage{color}
\usepackage[T1]{fontenc}
\usepackage[latin9]{inputenc}
\usepackage{array}
\usepackage{textcomp}
\usepackage{multirow}
\usepackage{soul}
\usepackage[short]{optidef} 
\usepackage{cleveref} 
\usepackage{bbm}
\usepackage{bm}
\usepackage{booktabs}
\usepackage{scalerel}
\usepackage[shortlabels]{enumitem}

\allowdisplaybreaks

\newtheorem{remark}{Remark}
\newtheorem{proposition}{Proposition}

\newcommand{\mc}[1]{\mathcal{#1}}

\newcommand{\mb}[1]{\mathbf{#1}}

\DeclareMathOperator*{\argmax}{arg\;max}

\newcommand{\bseq}{\begin{subequations}}
	\newcommand{\eseq}{\end{subequations}}
\newcommand{\baln}{\begin{align}}
\newcommand{\ealn}{\end{align}}
\newcommand{\balnd}{\begin{aligned}}
	\newcommand{\ealnd}{\end{aligned}}
\newcommand{\beq}{\begin{equation}}
\newcommand{\eeq}{\end{equation}}
\newcommand{\beqn}{\begin{eqnarray}}
\newcommand{\eeqn}{\end{eqnarray}}
\newcommand{\beqno}{\begin{eqnarray*}}
	\newcommand{\eeqno}{\end{eqnarray*}}
\newcommand{\bma}{\begin{displaymath}}
\newcommand{\ema}{\end{displaymath}}
\newcommand{\bnu}{\begin{enumerate}}
	\newcommand{\enu}{\end{enumerate}}
\newcommand{\bce}{\begin{center}}
	\newcommand{\ece}{\end{center}}
\newcommand{\btb}{\begin{tabular}}
	\newcommand{\etb}{\end{tabular}}
\newcommand{\ba}{\begin{array}}
	\newcommand{\ea}{\end{array}}

\ifCLASSINFOpdf
  
\else
 
\fi

\begin{document}
\bstctlcite{IEEEexample:BSTcontrol}
%
\title{A Hybrid Optimization and Deep RL Approach for Resource Allocation in Semi-GF NOMA Networks
}

\author{\IEEEauthorblockN{Duc-Dung Tran\textsuperscript{1}, Vu Nguyen Ha\textsuperscript{1}, Symeon Chatzinotas\textsuperscript{1},  and Ti Ti Nguyen\textsuperscript{2}
}
		\IEEEauthorblockA{
		\textsuperscript{1}\textit{Interdisciplinary Centre for Security, Reliability and Trust (SnT), University of Luxembourg, Luxembourg} \\
		\textsuperscript{2}\textit{Universit\'e du Qu\'ebec, Montr\'eal, Canada} \\
			Emails: \textsuperscript{1}\{duc.tran, vu-nguyen.ha, symeon.chatzinotas\}@uni.lu, \textsuperscript{2}titi.nguyen@emt.inrs.ca
			\vspace{-0.2cm}}
	}

\maketitle


\begin{abstract}
Semi-grant-free non-orthogonal multiple access (semi-GF NOMA) has emerged as a promising technology for the fifth-generation new radio (5G-NR) networks supporting the coexistence of a large number of random connections with various quality of service requirements. 
However, implementing a semi-GF NOMA mechanism in 5G-NR networks with heterogeneous services has raised several resource management problems relating to unpredictable interference caused by the GF access strategy. 
To cope with this challenge, the paper develops a novel hybrid optimization and multi-agent deep (HOMAD) reinforcement learning-based resource allocation design to maximize the energy efficiency (EE) of semi-GF NOMA 5G-NR systems.
In this design, a multi-agent deep Q network (MADQN) approach is employed to conduct the subchannel assignment (SA) among users.
While optimization-based methods are utilized to optimize the transmission power for every SA setting. 
In addition, a full MADQN scheme conducting both SA and power allocation is also considered for comparison purposes. 
Simulation results show that the HOMAD approach outperforms other benchmarks significantly in terms of the convergence time and average EE.

\end{abstract}


\IEEEpeerreviewmaketitle

\vspace{-0.1cm}

\section{Introduction}
The future wireless networks are expected to be capable of serving a tremendous number of devices requiring heterogeneous services, e.g., enhanced mobile broadband (eMBB), ultra-reliable low-latency communications (URLLC), and massive machine type communications (mMTC), together with different quality-of-service (QoS) demands \cite{LienCM17,TiTi_ComLet20}. 
In this context,
semi-GF NOMA has been considered as a promising solution for relieving the heavy accessing-process overhead in the dense systems \cite{ShahabCST20}. 
Following this strategy, 
the subchannels (SCs) are opened for mMTC users to access freely without waiting for receiving the admission granted, i.e., grant-free (GF) access, while the association process of other users having stringent QoS requirements (e.g., eMBB/URLLC users) are scheduled by the system controllers (such as base stations or access points, etc.), which is also called as grant-based (GB) access. 
In addition, the NOMA transmission can be exploited when there is more than one user accessing the same SC \cite{DungTran_SJ2020}.

However, the without-admission-control property of the GF strategy may result in a serious congestion problem in semi-GF NOMA systems when a tremendously large number of devices tries to access a limited number of SCs. 
Therefore, GF access needs to be carefully designed to mitigate this problem as well as guarantee the QoS requirements of both GB and GF users in semi-GF NOMA systems. 
Furthermore, in real-time systems, developing a dynamic resource allocation (RA) mechanism addressing the congestion problem and fulfilling the various QoS requirements from different services in semi-GF NOMA systems becomes more challenging. 
In recent years, reinforcement learning (RL) method has been applied to intelligently resolve the RA problem in communications \cite{ShahabCST20}. Its application to GF NOMA and semi-GF NOMA systems has been investigated in \cite{SilvaWCL20,DungTranGLOBECOM21,DungTranPIMRC21,DungTran_VTC21,DungTranVTC22,FayazTWC21,FayazArxiv22,LiuTCOM23,DungTranOJCOM23}. However, these works have not considered the 5G-NR systems with the coexistence of multiple services. Furthermore, most of them aimed to discretize the continuous power variable to ease the learning process which may result in performance loss.

Regarding the drawback of the existing works, this paper develops two novel learning-based resource allocation designs maximizing EE while guaranteeing heterogeneous requirements relating to various communication services in semi-GF NOMA 5G-NR systems. Both of these proposed algorithms exploit the multi-agent deep RL method where the mMTC users are considered as agents that learn and optimize its SC and transmission power selection.
The first algorithm, namely full multi-agent deep Q-network (Full-MAD), aims to set both SC assignment and power allocation (PA) as the action for the learning process, where the transmission power is quantized into a number of discrete levels.
On a different method, the second algorithm, namely HOMAD, only considers the SC selection as the action model. In this learning-based solution, the transmission power corresponding to each SC setting can be determined by some efficient optimization-based analysis results.
By doing so, the action space size is significantly degraded and the hybrid method can take advantage of both deep Q-network (DQN) and optimization-based approaches to gain better learning performance.
The simulation results are then demonstrated to evaluate the performance of our proposed mechanisms in terms of convergence time and the system EE.


\section{System Model}
\label{sec_system_model}

\begin{figure}[!t]
	\centering
	\includegraphics[width=7cm,height=3cm]{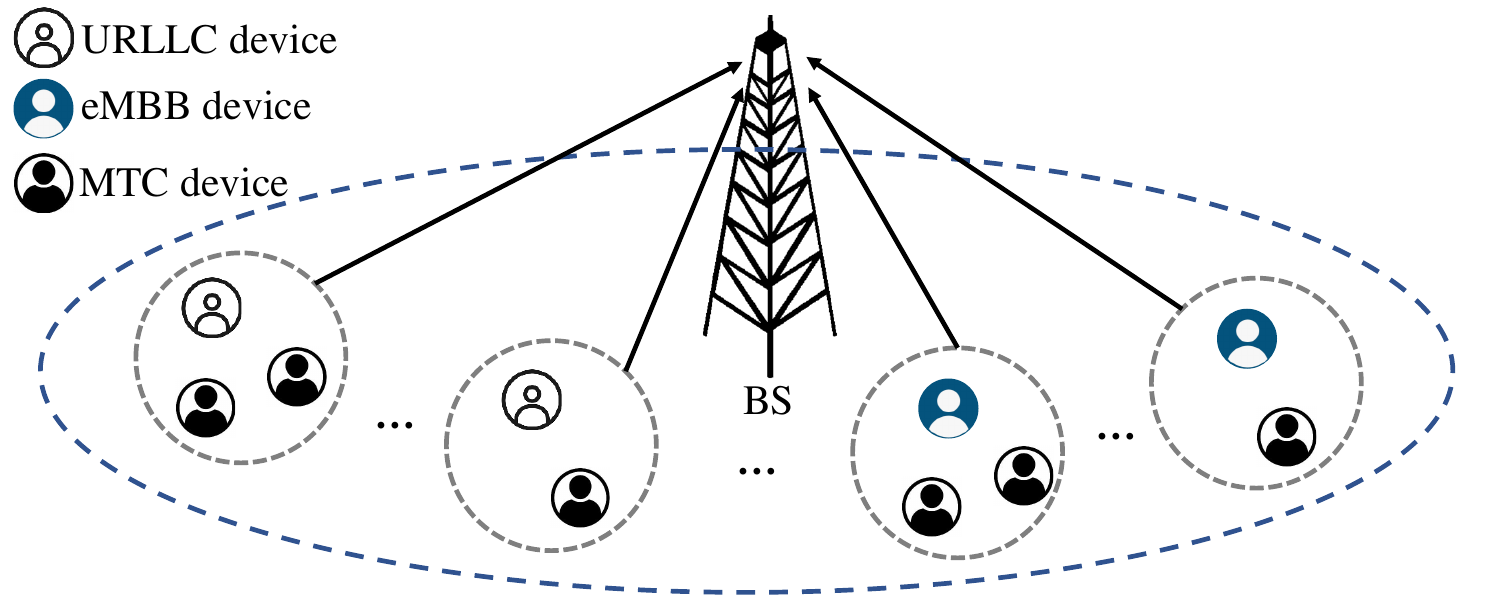}
	\caption{Illustration of an uplink semi-GF NOMA 5G-NR network.}
	\label{fig_model}
	\vspace{-15pt}
\end{figure}

We investigate an uplink semi-GF NOMA 5G-NR network as shown in Fig. \ref{fig_model}. The network consists of one BS located at the center of the cell with a radius of $r$ (m) and a number of users randomly distributed in this cell requiring different services including eMBB/mMTC/URLLC. 
Let $\mc{M}_{\sf{U}}$, $\mc{M}_{\sf{E}}$ and $\mc{M}_{\sf{M}}$ be the sets of URLLC, eMBB, and mMTC devices, whose cardinalities are $M_{\sf{U}}$,  $M_{\sf{E}}$ and $M_{\sf{M}}$, respectively. For convenience, we also denote the set of all users as $\mc{M}= \mc{M}_{\sf{U}} \cup \mc{M}_{\sf{E}} \cup \mc{M}_{\sf{M}}$ and $M = M_{\sf{U}} + M_{\sf{E}} + M_{\sf{M}}$. 
To serve these users, a total bandwidth of $W$~(Hz) is assumed in the system, which is divided into $K$ SCs.
Let $\mc{K}$ be the set of all SCs.
Furthermore, to guarantee heterogeneous requirements of different services, a semi-GF NOMA transmission scheme is considered for the communication process.

\vspace{-0.1cm}

\subsection{Uplink Semi-GF NOMA 5G-NR Transmission Strategy} \label{sec_sgf_noma_strategy}
\subsubsection{5G-NR Numerology}
Following the 5G-NR standard which introduces various \textit{``numerologies''}, physical-resource-block (PRB) or SC types, supporting different communication requirements, the bandwidth of SC in 5G-NR schemes is defined as $2^\nu$ times SC's bandwidth in 4G systems (i.e., 180 kHz), where $\nu \in \{0; 1; 2; 3; 4\}$ is the numerology index \cite{LienCM17,VuHaNL20}. 
Herein, PRBs with high SC spacing are arranged for URLLC services while traffic flows from the eMBB
service can adopt a numerology with the smaller SC spacing \cite{LienCM17}.
Therefore, this paper focuses on an SC setting that the whole bandwidth is divided into two sets of SCs, $\mc{K}_{\sf{U}}$ and $\mc{K}_{\sf{E}}$.
Particularly, $\mc{K}_{\sf{U}}$ represents the set of SCs serving URLLC users with numerology $\nu_{\sf{U}}$ while $\mc{K}_{\sf{E}}$ is the set of eMMB-service SCs with numerology $\nu_{\sf{E}}$. 
Herein, $\mc{K}_{\sf{U}} \cup \mc{K}_{\sf{E}} = \mc{K}$.
One assumes that $\nu_{\sf{E}} < \nu_{\sf{U}}$ and denotes $W_{\sf{E}} = 2^{\nu_{\sf{E}}} \times 180$ (kHz) and $W_{\sf{U}} = 2^{\nu_{\sf{U}}} \times 180$ (kHz) as the bandwidth of SCs corresponding to eMBB and URLLC services, respectively.
\subsubsection{Semi-Grant-Free Radio Access Strategy}
In this system, the radio access of eMBB and URLLC users is
managed by BS under the GB access scheme due to their requirements for high reliability, latency, and achievable rate. 
Specifically, each of these users is granted several distinct SCs for its transmission. 
In contrast, the mMTC users can access the network based on GF access method to improve connection density due to the massive access requirement of mMTC service. 
Herein, the mMTC users can access any SCs freely without a scheduling process to increase the access rate and the number of active mMTC users. 
In this context, many mMTC users can access the same SC; furthermore, they can use the SCs which are already granted to the eMBB and URLLC users. 

 Considering the transmission over SC $k$ ($k \in \mathcal{K}$), we denote $b_z^{(k)}(t)$ ($z \in \mc{M}$) as a binary SC allocation variable at time-slot (TS) $t$, where $b_z^{(k)}(t) = 1$ if device $z$ occupies SC $k$ and $b_z^{(k)}(t) = 0$ otherwise.
In our scheme, we assume the orthogonal SC scheduled for URLLC/eMBB services and one-SC freely access strategy for mMTC users where each mMTC device can select only one arbitrary SC for its transmission. This assumption yields the following conditions,
 \beqn
 (C1): &\quad \sum_{z \in \mc{M}_{\sf{U}}\cup\mc{M}_{\sf{E}}} b_z^{(k)}(t) \leq 1, \quad \forall k \in \mc{K}. \\
  (C2): &\quad \sum_{k \in \mc{K}} b^{(k)}_z(t) = 1, \quad \forall z \in \mc{M}_{\sf{M}}.
 \eeqn
In addition, the set of devices occupying SC $k$ in TS $t$ can be described as $\mathcal{Z}^{(k)}(t) = \{z | b_z^{(k)} = 1, z \in \mc{M}\}$.
\subsubsection{NOMA Transmission Mechanism} \label{sec_noma_mechanism}
In uplink NOMA, the decoding order of the multi-user data stream is affected by various different factors. Specifically, a decoding order can be formulated based on channel gain conditions \cite{HanTVT19}, received power levels \cite{SilvaWCL20}, or QoS constraints of users \cite{DingCOML20,DungTran_TVT2021}.
In this paper, the messages of the users over each SC can be decoded at the BS as follows:
\begin{itemize}
\item Due to strict QoS requirements on reliability and latency, the URLLC user's signal needs to be decoded first.
\item The symbols belonging to eMBB and mMTC users will be decoded in the order of the corresponding channel gains. In particular, the user having the higher channel gain will be decoded earlier at the BS.
\item After decoding the message of a user with higher channel gain, the BS removes this component from its observation to decode the remaining users' messages by using the successive interference cancellation (SIC) technique.
\end{itemize}
Without loss of generality, one assumes there are $Z^k$ users accessing SC $k$ in TS, then they are arranged in the decoding order discussed above as $\mathcal{Z}^{(k)}(t) = \left\{z^{(k)}_1,...,z^{(k)}_{Z^k}\right\}$.
Accordingly, the received signal-to-interference-plus-noise ratio (SINR) of user $z^{(k)}_{\ell}$ is expressed as
\beq \label{SINR_ordering}
\gamma^{(k)}_{z^{(k)}_{\ell}}(t) = {\mc{Y}_{z_l^{(k)}}^{(k)}(t) }/{\big( \sum_{j>\ell} \mc{Y}_{z_j^{(k)}}^{(k)}(t) + \sigma_k^2 \big)},
\eeq
where $\mc{Y}_z^{(k)}(t) = P_z^{(k)}(t) g_z^{(k)}(t)$ is the power of signal due to user $z$'s data  over SC $k$ in TS $t$; $P_z^{(k)}(t)$ is the transmission power of user $z$ over SC $k$, in which $P_z^{(k)}(t) = 0$ if $b_z^{(k)}(t) = 0$ and $P_z^{(k)}(t) \ne 0$, otherwise; $g^{(k)}_z(t)$ denote the corresponding channel gain and $\sigma_k^2$ represents the noise power over SC $k$.
\vspace{-0.1cm}

\subsection{Achievable Rate of Users}
\subsubsection{URLLC Communication}

Regarding the transmission of URLLC user $u$ over SC $k$ in $\mc{K}_{\sf{U}}$, which happens when $b^{(k)}_u = 1$. 
Based on the NOMA transmission mechanism given in Section~\ref{sec_sgf_noma_strategy}, one must have $u \equiv z^{(k)}_1$. Moreover, the SINR of URLLC device $u$ over SC $k$ is expressed as
\begin{equation}
    \gamma_u^{(k)}(t) = {\mc{Y}_u^{(k)}(t)}/{\left( \mc{I}_u^{(k)}(t) + \sigma_u^2 \right)},
    \label{eq_SINR_uk}
\end{equation}
where $\mc{I}_u^{(k)}(t)=\sum_{j=2}^{Z^k} \mc{Y}_{z_j^{(k)}}^{(k)}(t)$ represents the interference caused by mMTC users over SC $k$. Furthermore, bandwidth of SC $k$ in $\mc{K}_{\sf{U}}$ is $W_{\sf{U}}$ and $\sigma_u^2 = F N_0 W_{\sf{U}}$ denotes the noise power, where $F$ is the noise figure, $N_0$ is the noise power spectral density (PSD). Accordingly, the achievable rate of URLLC user $u$ over SC $k$ in finite blocklength regime for a quasi-static flat fading channel can be approximated as \cite{Sun2019TWC}
\begin{equation}
    \begin{split}
        {R_u^{(k)}}(t) &= {W_{\sf{U}}}[ {{{\log }_2}( 1 + \gamma_u^{(k)}(t) )}  - \Phi_u^{(k)}(t)  ],
    \end{split}
    \label{eq_Rukt}
\end{equation}
where $\Phi_u^{(k)}(t) = \scaleobj{.8}{\sqrt {\frac{{{V^{(k)}_u}(t)}}{{{D_{u}}{W_{\sf{U}}}}}} \frac{{{Q^{ - 1}}\left( \varepsilon_u \right)}}{{\ln 2}}}$, $V^{(k)}_u(t) = 1 - \scaleobj{.8}{\left( 1 + \gamma^{(k)}_u(t) \right)^{-2}} \approx 1$ \cite{Sun2019TWC} is the channel dispersion, $\varepsilon_{u}$ is the decoding error probability (DEP) which can be used to evaluate the transmission reliability, $D_{u}$ is the transmission latency threshold, and $Q^{-1}(x)$ is the inverse of the Gaussian Q-function. Here, we define a data-rate demand for URLLC $u$ to satisfy the URLLC requirements (i.e., $\varepsilon_u$ and $D_{u}$) when transmitting one packet over one SC in each TS as $R_{u}^{\sf{tar}} = {W_{\sf{U}}}\left[ {{\log_2}\left( 1 + \gamma_{u}^{\sf{tar}} \right)}  - \Phi_{u}^{\sf{tar}} \right]$,
where $\gamma_{u}^{\sf{tar}} = 2^{\frac{n_b}{D_{u}W_{\sf{U}}} + \frac{Q^{-1}\left( \varepsilon_{u} \right)}{\ln{2} \sqrt{D_{max}W_{\sf{U}}}} } - 1$ is the target SNR for user $u$ \cite{Sun2019TWC}, $n_u$ is the packet size, and $\Phi_{u}^{\sf{tar}}$ is defined similarly as in \eqref{eq_Rukt}. This demand yields the following constraints,
\beq
 (C3): \quad  b_u^{(k)}(t){R_u^{(k)}}(t) \geq R_{u}^{\sf{tar}}, \quad \forall k \in \mc{K}.
\eeq

\vspace{-0.1cm}

\subsubsection{eMBB Communication}
Assume that eMBB user $e$ access SC $k$ in $\mc{K}_{\sf{E}}$ which implies $b^{(k)}_e = 1$. Due to its order in the NOMA-based decoding process, its SINR denoted as $\gamma_e^{(k)}\left( t \right)$, can be defined as in \eqref{SINR_ordering} with noting that $\sigma_k^2 = F N_0 W_{\sf{E}}$. Then, the achievable rate of eMBB device $e$ is given by
\begin{equation}
    R_e^{(k)}(t) = W_{\sf{E}} \log_2 \left( 1 + \gamma_e^{(k)}(t) \right).
    \label{eq_Rekt}
\end{equation}
Herein, one addresses a predetermined target transmission rate, $R_e^{\sf{tar}}$, for each eMBB  user $e$ in every TS as
\begin{equation}
(C4): \quad \scaleobj{.8}{\sum_{k \in \mathcal{K}}} b_e^{(k)}(t) R_e^{(k)}(t) \ge R_e^{\sf{tar}}, \quad \forall e \in \mc{M}_{\sf{E}}.
\end{equation}

\vspace{-0.1cm}

\subsubsection{mMTC Communication}
Based on the NOMA transmission strategy mentioned earlier in Section \ref{sec_sgf_noma_strategy}, mMTC devices can select a free SC or the one occupied by either URLLC or eMBB device. 
When $b^{(k)}_m(t) = 1$, mMTC user $m$ utilize SC $k$ in TS $t$. 
In such case, the SINR of this device, denoted as $\gamma_m^{(k)}(t)$, can be calculated as in \eqref{SINR_ordering} with noting that $\sigma_k^2 = F N_0 W_k$ where $W_k = W_{\sf{E}}$ if $k \in \mc{K}_{\sf{E}}$, and $W_k = W_{\sf{U}}$, otherwise.
Similar to URLLC devices, the achievable rate of mMTC device $m$ is given by $R_m^{(k)}(t) = W_k \left[ \log_2 ( 1 + \gamma_m^{(k)}(t) ) - \Phi_m^{(k)}(t) \right]$, where $\Phi_m^{(k)}(t) = \scaleobj{.8}{\sqrt {\frac{{{V^{(k)}_m}(t)}}{{{D_{m}}{W_{\sf{k}}}}}} \frac{{{Q^{ - 1}}\left( \varepsilon_m \right)}}{{\ln 2}}}$, $V^{(k)}_m(t) = 1 - \scaleobj{.8}{\left( 1 + \gamma^{(k)}_m(t) \right)^{-2}} \approx 1$ \cite{Sun2019TWC}.
Furthermore, the target SNR of mMTC device $m$ can be defined as $\gamma_{m}^{\sf{tar}} = 2^{\frac{n_m}{D_{m}W_{k}} + \frac{Q^{-1}\left( \varepsilon_{m} \right)}{\ln{2} \sqrt{D_{m}W_{k}} }} - 1$, where $n_m$, $D_m$, and $\varepsilon_m$ denote the packet size, transmission latency, and DEP of mMTC device $m$.
Then, the SINR of mMTC user should be greater than a threshold for successful decoding, i.e.,
\beq
(C5): \quad b_m^{(k)}(t)\gamma_m^{(k)}(t) \geq \gamma_m^{\sf{tar}}, \quad \forall m \in \mc{M}_{\sf{M}}.
\eeq

\vspace{-0.1cm}

\subsection{Energy Efficiency Maximization Problem}
In this paper, we aim to design an effective SC and power allocation strategy to maximize the network EE while guaranteeing the different requirements of all services. To do so, we first define an EE factor as follows:
\begin{equation}
    \zeta(t) = {R^{\sf{tot}}(t)}/{(P^{\sf{Tx}}(t) + M P_{\sf{c}})},
    \label{eq_EEt}
\end{equation}
where $R^{\sf{tot}}(t) = \sum_{k \in \mc{K}} \sum_{z \in \mc{M}} b_z^{(k)}(t) R_z^{(k)}(t)$, $P^{\sf{Tx}}(t) = \sum_{k \in \mc{K}} \sum_{z \in \mc{M}} P_z^{(k)}(t)$, and $P_{\sf{c}}$ denotes the circuit power consumption.
Then, the design problem can be formulated as
\begin{subequations} \label{max_ee}
	\begin{eqnarray} 
			\hspace{-1cm}\underset{\mb{b},\mb{P}}{\max}  \; \mathbb{E}_t\left[ \zeta(t) \right] &\text{ s.t.} & \text{ constraints } (C1)-(C5), \\
			\hspace{-1cm} && \hspace{-1cm} (C6): \scaleobj{.8}{\sum_{k \in \mc{K}}} P_z^{(k)}(t) \le P_z^{\sf{max}}, \; \forall (z,t), \label{c8}
		\end{eqnarray}
	\end{subequations}
where $\mb{b}$ and $\mb{P}$ denote the SC assignment and power control strategies, respectively; and constraint $(C6)$ stands for the power budget of devices.

\section{Two Proposed Multi-Agent Deep RL Solutions}

\subsection{Full multi-agent DQN Approach} \label{sec_FMADQN_alg}
 A full multi-agent DQN approach, named Full-MAD, is first studied in this section.
 Herein, all mMTC users are considered agents.
 Employing a multi-agent deep RL mechanism, they separately learn and define optimal policies for selecting SC and PA. 
 In addition, the multi-level quantization strategy is exploited to deal with the continuous characteristic of power variables in the similar approach introduced in \cite{SilvaWCL20,FayazTWC21}.
 Herein, the power is quantized into $L$ levels to build the action sets for the RL process. 
 Particularly, the state, action, and reward of each agent ( e.g., $m \in \mc{M}_{\sf{M}}$) in TS $t$ are defined as follows. The state of agent $m$ is defined as
\beq \label{eq_statem}
s_m(t) = \left\{ g_m^{(1)}(t), \dots, g_m^{(K)}(t), \mb{c}_m(t-1) \right\},
\eeq
where $\mb{c}_m(t-1) = \{ b_m^{(1)}(t), \dots, b_m^{(K)}(t), P_m^{(1)}(t), \dots, P_m^{(K)} \}$ is the SC and transmission power level (TPL) selection status of agent $m$ in TS $t-1$.
Additionally, one defines the action of agent $m$ as its SC and TPL selection, expressed as
\beq \label{eq_actionm}
a_m(t) = \{ 1, \dots, kl, \dots, KL \},
\eeq
where $a_m(t) = kl$ indicates that agent $m$ select SC $k$ and the $l$-th TPL in TS $t$. Let $\mc{A}_m$ be the set of all actions of mMTC user $m$, we have $\left| \mc{A}_m \right| = KL$.
For action selection strategy, the $\epsilon$-greedy policy can be exploited where the random action is taken with the probability of $\epsilon$, and the action with the highest Q-value, i.e., $a_m^{max} = {\argmax}_{a \in \mc{A}_m}\left\{Q_m \left( s_m(t), a; \bm{\theta}_m \right)\right\}$ is employed for the remaining probability. Herein, $Q_m\left( s_m(t), a_m(t); \bm{\theta}_m \right)$ is the Q-value corresponding to action $a_m(t)$.
Regarding the EE factor, we can define the reward function in TS $t$  as
\beq \label{eq_reward}
r(t) = \left\{ \begin{array}{*{20}{c}}
                        {\zeta(t),} & \text{if all constraints are satisfied,}\\
                        {0,} & \text{otherwise.}
                        \end{array} \right.
\eeq
Based on the actions and rewards obtained from trials, each agent builds its own DQN model consisting of two deep neural networks (DNNs), namely online and target networks corresponding to weight vectors $\bm{\theta}_m$ and $\bm{\theta'}_m$, respectively. 
Herein, the online network is used to select an action. 
Meanwhile, the target network is applied to evaluate the online network-based action. Thus, the objective is to reduce the loss function as \cite{FayazTWC21}
\beq \label{eq_loss}
\hat{L}(\bm{\theta}_m) = \left[ y_m(t) - Q_m\left( s_m(t), a_m(t); \bm{\theta}_m  \right) \right]^2,
\eeq
where $y_m(t)$ denotes the target Q-value determined by the target network as $y_m(t) = r(t) + \underset{a \in \mc{A}_m}{\max}{Q_m \left( s_m(t+1), a; \bm{\theta'}_m \right)}$.
Given the MDP principle and DQN model of each agent mentioned above, the Full-MAD learning-based solution approach is summarized in Algorithm \ref{alg_fmadqn_homadqn}.

\begin{algorithm}[!t]
\caption{\footnotesize \textsc{Multi-Agent DRL-based Energy Efficiency Maximization Algorithm}}
\label{alg_fmadqn_homadqn}
\footnotesize
\begin{algorithmic}[1]
\STATE Initialize the weight vectors of the online and target networks, i.e., $\bm{\theta}_m$ and $\bm{\theta'}_m$, $\forall m \in \mc{M}_{\sf{M}}$.
\FOR{$i = 1, \dots, E_p$}
    \STATE Initialize the state $s_m(t)$, $\forall m$.
    \FOR{$t = 1, \dots, T$}
        \STATE All agents take actions as \begin{itemize}
            \item \textbf{For Full-MAD approach:} \textit{Both SC and power-level selection} (as in \eqref{eq_actionm})  following the $\epsilon$-greedy policy.
            \item \textbf{For HOMAD approach:}  \textit{Only SC selection} (as in \eqref{eq_actionm_homad}) following the $\epsilon$-greedy policy. \textit{Optimizing transmission power} as in Section~\ref{Sec:HOMAD}.
        \end{itemize}  
        \STATE All agents observe the reward in \eqref{eq_reward} and move to the next states.
        \FOR{$m = 1, \dots, M_{\sf{M}}$}
            \STATE Store an experience tuple $\left( s_m(t), a_m(t), r(t), s_m(t+1) \right)$ to the memory of agent $m$.
            \STATE Randomly sample a mini-batch of experiences from the memory to train the online network.
            \STATE Update $\bm{\theta}_m$ by using gradient descent to minimize the loss function in \eqref{eq_loss}.
            \STATE Update $\bm{\theta'}_m$ as $\bm{\theta'}_m = \bm{\theta}_m$ after every $B$ TSs.
        \ENDFOR
    \ENDFOR
\ENDFOR
\end{algorithmic}
\normalsize
\end{algorithm}

\subsection{Hybrid Optimization and Multi-Agent DQN Approach} \label{Sec:HOMAD}

In this section, we developed the HOMAD approach to speed up the learning process and eliminate the power-quantization loss. Specifically, the HOMAD mechanism employs a MADQN-based method for SC selection according to which the transmission power is optimized efficiently. In this approach, the states and rewards are defined similarly to those presented in the Full-MAD method while the action is simplified to SC selection as
\beq \label{eq_actionm_homad}
a_m(t) = \{ 1, \dots, k, \dots, K\}.
\eeq
Once the action is taken, the transmission power is optimized as follows.
Firstly, employing the Dinkelbach algorithm \cite{VuHaTGCN20,VuHa_ICC18}, we aim to solve problem \eqref{max_ee} by iteratively solving, 
\beq \label{sub-tract-prob}
\hspace{-10pt} \underset{\mb{p}(t)}{\max} \; R^{\sf{tot}}(t) - \zeta  P^{\sf{Tx}}(t) \text{ s.t. $(C1)-(C6)$,}
\eeq
and adjusting $\zeta$ until an optimal $\zeta^{\star} \geq 0$ satisfying $R^{\sf{tot}}(t) = \zeta^{\star} \left(P^{\sf{Tx}}(t) + M P_{\sf{c}} \right)$ is found.
To cope with \eqref{sub-tract-prob}, let's regard the following remark based on which we propose efficient approaches to optimize the power for each SC setting.

\begin{remark} \label{rmk02}
The formula given in \eqref{SINR_ordering} demonstrates that there is no interference suffering the decoding process due to user $z^{(k)}_{Z^k}$. Moreover, once the power of all users in set $\left \{ \ell +1, ..., z^{(k)}_{Z^k}\right \}$ is defined, the transmission power of user $z^{(k)}_{\ell}$, i.e., $P^{(k)}_{z^{(k)}_{\ell}}$, can be optimized without coupling to other users. 
\end{remark}


\subsubsection{Power Allocation for mMTC Service} \label{Sec:mMTC_op}
Once user $z^{(k)}_{\ell}$ is an mMTC user, $P^{(k)}_{\ell}$ can be optimized for given $\lbrace P^{(k)}_{m}\rbrace _{m=\ell+1}^{Z_k}$ by solving the following sub-problem.
\beq \label{pa_mMTC}
\underset{p_{\ell}}{\max} \; R_m^{(k)}(t) - \zeta p_{\ell} \text{ s.t. } {{\gamma}_{z^{(k)}_{\ell}}^{\sf{tar}}}/{A^{(k)}_{\ell}} \leq p_{\ell} \leq P^{\sf{max}}_{z^{(k)}_{\ell}},
\eeq
where $A^{(k)}_{\ell} \! = \! {g_{z^{(k)}_{\ell}}^{(k)}\!(t) }/{(\sum_{j>\ell}g_{z^{(k)}_{j}}^{(k)}(t) P^{(k)}_{z^{(k)}_{j}}\! +\! \sigma_k^2)}$.

\begin{proposition} The solution of problem \eqref{pa_mMTC} is given as,
 \beq \label{op_pw_mMTC}
 P^{(k)\star}_{\ell}\!\! = \! \min ( \max ( {W_k}/{(\zeta \ln{2})} \! - \! {1}\!/\!{A^{(k)}_{\ell}}\!\! , \! {{\gamma}_{z^{(k)}_{\ell}}^{\sf{tar}}}\!/\!{A^{(k)}_{\ell}}\!),\! P^{\sf{max}}_{z^{(k)}_{\ell}}). \!\!\!
 \eeq
 \end{proposition}
 \begin{IEEEproof}
 The proof is described simply as follows. 
 As can be seen, problem \eqref{pa_mMTC} is convex due to the concave objective function and the convex feasible set. Then, the optimal solution can be obtained by setting the derivative of the objective function to zero and solving it with the feasible set \cite{VuHa_TWC21}. 
 \end{IEEEproof}
\subsubsection{Power Allocation for eMBB Service} \label{Sec:eMBB_op}
Assume that user $e$ is assigned $n$ SCs named as $\{k^e_1,...,k^e_n \} \subset \mc{K}_{\sf{E}}$, and it is denoted as user $z^{(k^e_j)}_{\ell_j}$ over SC $k^e_j$ ($j = 1,...,n$).
Then, problem \eqref{sub-tract-prob} can be decomposed for user $e$ as
\beq \label{eMBB_prob}
			\scaleobj{.9}{\underset{\mb{P}^e}{\max} \sum\nolimits_{j=1}^n C_j^e - \zeta p^e_j
			\text{ s.t.} \scaleobj{.9}{\sum\nolimits_{j=1}^n} C_j^e \geq R^{\sf{tar}}_e, \scaleobj{.9}{\sum\nolimits_{j=1}^n} p^e_j \leq P_e^{\sf{max}}}, 
\eeq
where $C_j^e = W_{\sf{E}}\log_2\left(1+A^e_j p^e_j\right)$, $\mb{P}^e = [p^e_1,...,p^e_n]$, $p^e_j$ is the transmission power variable of eMBB user $e$ over SC $k^e_j$, and $A^e_j$ is defined similarly as in \eqref{pa_mMTC}. Since problem \eqref{eMBB_prob} is convex, its solution can be obtained by using the duality method.
In particular, the Lagrangian of \eqref{eMBB_prob} is described as
$\mc{L}(\mb{P}^e,\mu, \nu) = \sum_{j=1}^n \left[ (1+\mu) C_j^e - (\zeta + \nu) p^e_j \right] -\mu R^{\sf{tar}}_e + \nu P_e^{\sf{max}}$,
where $\mu$ and $\nu$ are the Lagrangian multipliers corresponding to the constrains of  \eqref{eMBB_prob}.
Then, the dual function is defined as
${\sf{g}}(\mu,\nu) = \underset{\mb{P}^e}{\max} \; \mc{L}(\mb{P}^e,\mu, \nu)$. 
\begin{proposition}
The solution of dual function is defined as
\beq
p^e_j = \max\left({(1+\mu)W_{\sf{E}}}/{[(\nu+\zeta)\ln{2}]} -{1}/{A^e_j},0\right).
\eeq
\end{proposition}
\begin{IEEEproof}
The proof of this proposition can be obtained easily by solving the equation $\partial \mc{L}(\mb{P}^e,\mu, \nu)/\partial p^e_j = 0$. 
\end{IEEEproof}
The dual problem can be rewritten as
$\underset{\mu,\nu}{\max}\;{\sf{g}}(\mu,\nu)$ $\text{s.t. } \mu, \nu \geq 0$.
Since problem \eqref{eMBB_prob} is convex and the dual gap is zero, the optimal solution of the dual problem can be found by iteratively updating the dual variables $\mu$ and $\nu$ as $
\mu^{[v +1]} \!\! = \!\!  \left[ \mu^{[v]} \!\! - \! \delta^{[v]} \! (\sum_{j=1}^n \!\! C^e_j \!\! - \!\! R^{\sf{tar}}_e)\right]^+$ and $\nu^{[v +1]} \!\! =\! \! \left[ \nu^{[v]} \!\! + \!\! \delta^{[v]} \! (\sum_{j=1}^n \! p^e_j \!\! - \!\! P_e^{\sf{max}})\right]^+$,
where the suffix $[v]$ represents the iteration index, $\delta_{[v]}$ is the step size.
This sub-gradient method guarantees the convergence if $\delta_{[v]} \overset{v \rightarrow \infty}{\longrightarrow} 0$ \cite{BoydBook}.

\subsubsection{Power Allocation for URLLC Service} \label{Sec:URLLC_op}
Similar to the previous section, one assumes that there are $l$ SCs assigned to URLLC user $u$, namely $\{k^u_1,...,k^u_l\} \subset \mc{K}_{\sf{U}}$. 
Then, if the power of all mMTC users on SCs $\{k^u_1,...,k^u_l\}$ are determined, the power transmission over all SCs can be determined by solving the following problem
\begin{subequations} \label{URLLC_prob}
	\begin{eqnarray} 
			\hspace{-1cm}&\underset{\mb{P}^u}{\max} \sum\nolimits_{j=1}^l \left( C^u_j - \Phi_j^u\right) - \zeta p^u_j 
			\text{ s.t.} &  p^u_j \geq \gamma_u^{\sf{tar}}/A^u_j, \; \forall j, \label{cnt_URLLC01} \\
			\hspace{-1cm} && \scaleobj{.95}{\sum\nolimits_{j=1}^l} p^u_j \leq P_u^{\sf{max}}, \label{cnt_URLLC02}
	\end{eqnarray}
\end{subequations}
where $C^u_j\!\! = \!\! W_{\sf{U}}\log_2\left(1+A^u_j p^u_j\right)$, $\Phi_j^u = \sqrt {\frac{V_j^u}{D_{u} W_{\sf{U}}}} \frac{Q^{-1}\left( \varepsilon_j^u \right)}{\ln 2}$, $V_j^u = 1 - \left( 1 + A^u_j p^u_j \right)^{-2} \approx 1$ \cite{Sun2019TWC}, $\mb{P}^u = [p^u_1,...,p^e_l]$ and $p^u_j$ denotes the transmission power variable corresponding to URLLC user $u$ over SC $k^u_j$.
Similar to the approach employed for solving problem \eqref{eMBB_prob}, the transmission power of URLLC users can be determined in the following proposition.
\begin{proposition} The transmission power of URLLC users
$u$ over SCs $\{k^u_1,...,k^u_l\}$, can be defined as
\beq
p^u_j = \max\left({W_{\sf{U}}}/{\big[ (\theta+\zeta)\ln{2} \big]} -{1}/{A^u_j},{\bar{\gamma}_u}/{A^u_j}\right), \forall j,
\eeq
where $\theta$ is iteratively updated as $ 
\theta^{[v +1]} \!\!\!\! = \!\!\!\! \left[ \theta^{[v]} + \delta^{[v]} (\sum_{j=1}^l p^u_j - P_u^{\sf{max}})\right]^+$.
\end{proposition}
\begin{IEEEproof}
The proof can be obtained by employing the similar duality method presented in Section~\ref{Sec:eMBB_op}. 
\end{IEEEproof}
In summary, the proposed energy-efficiency power allocation algorithm is described in Algorithm~\ref{P2_alg:1}.


\begin{algorithm}[!t]
\caption{\footnotesize \textsc{Energy-Efficiency Power Allocation Algorithm}}
\label{P2_alg:1}
\footnotesize
\begin{algorithmic}[1]
\STATE Initialize $\zeta^{(0)}=0$, set $q=0$, and choose predetermined tolerate $\tau$.
\REPEAT 
\STATE The power allocation can be optimized in a parallel manner over all SCs for mMTC users, but the process over an SC involved in an eMBB user process can stop and then continue when the power of that eMBB user is updated. The process is described as
\begin{enumerate}[a.]
    \item \textbf{The power of every mMTC user} are defined as in \eqref{op_pw_mMTC}.
    \item \textbf{The power of every eMBB user} is optimized as described in Section~\ref{Sec:eMBB_op} when all mMTC ordered before it over the corresponding SCs having their transmission power optimized.
    \item \textbf{The power of every URLLC user} is optimized as in Section~\ref{Sec:URLLC_op} when all mMTC users have their power transmission determined.
\end{enumerate}
\STATE Update $\zeta^{(q+1)}=\frac{R^{\sf{tot}}(t)}{P^{\sf{Tx}}(t) + M P_{\sf{c}}}$.
\STATE Set $q:=q+1$.
\UNTIL $\vert \zeta^{(q)} - \zeta^{(q-1)} \vert \leq \tau$.
\end{algorithmic}
\normalsize
\end{algorithm}

\subsubsection{Proposed HOMAD Algorithm}
As mentioned above, the HOMAD approach allows agents to select SCs by using a MADQN scheme similar to Full-MAD approach whereas the transmission power for each SC setting is optimized by applying Algorithm \ref{P2_alg:1}. The summary of this algorithm is also provided in Algorithm~\ref{alg_fmadqn_homadqn} with \textbf{``HOMAD'' remark} in \textbf{Step 5}.


\section{Simulation Results}
\begin{table}[!t]
    \caption{Experimental Parameters}
    \vspace{-5pt}
    \scriptsize
    \centering
    \begin{tabular}{|l|l|}
        \hline
        \textbf{Parameters} & \textbf{Value} \\
        \hline
        Cell radius ($r$) & 500 m \\
        \hline
        Channel model & Rician \\
        \hline
        eMBB, URLLC numerology indices ($\nu_{\sf{E}},\nu_{\sf{U}}$) & 1, 4 \\
        \hline
        eMBB Data-rate demand $\left( R_e^{\sf{tar}} \right)$ & \{2; 4; 6; 8\} bps/Hz \\
        \hline
        Latency threshold $\left( D_u = D_m = D_{max} \right)$ & \{2; 1; 0.5; 0.4\} ms \\
        \hline
        Reliability threshold $\left( \varepsilon_{u} = \varepsilon_m = \varepsilon_{th} = 10^{-x} \right)$ & $x = \left\{2; 4; 5, 6, 7 \right\}$ \\
        \hline
        Maximum transmission power & 23 dBm \\
        \hline
        Circuit power consumption ($P_c$) & 0.05 W \\
        \hline
        Noise figure and PSD ($F$ and $N_0$) & $6$ dB and -174 dBm/Hz \\
        \hline
        Packet length ($n_u = n_m = n_b$) & 32 bytes \\
        \hline
        Number of hidden layers, neurons per hidden layers & 3, \{256, 128, 64\} \\
        \hline
        Learning rate ($\alpha$) and discount factor ($\gamma$) & 0.001 and 0.9 \\
        \hline
        Optimizer & Adam \\
        \hline
    \end{tabular}
    \label{tab:experimental_para}
    \normalsize
    \vspace{-10pt}
\end{table}
This section provides the simulation results to evaluate our proposed algorithms' performance. The simulations were performed on a PC equipped an Intel Xeon W-11855M CPU with 3.2 GHz frequency, 64-GB RAM, and 64-bit Windows 10 operating system. The DQN model consists of three fully-connected hidden layers including 256, 128, and 64 neurons. The experimental parameters are provided in Table \ref{tab:experimental_para}.
\begin{figure}[!t]
	\centering
	\includegraphics[width=8cm,height=5cm]{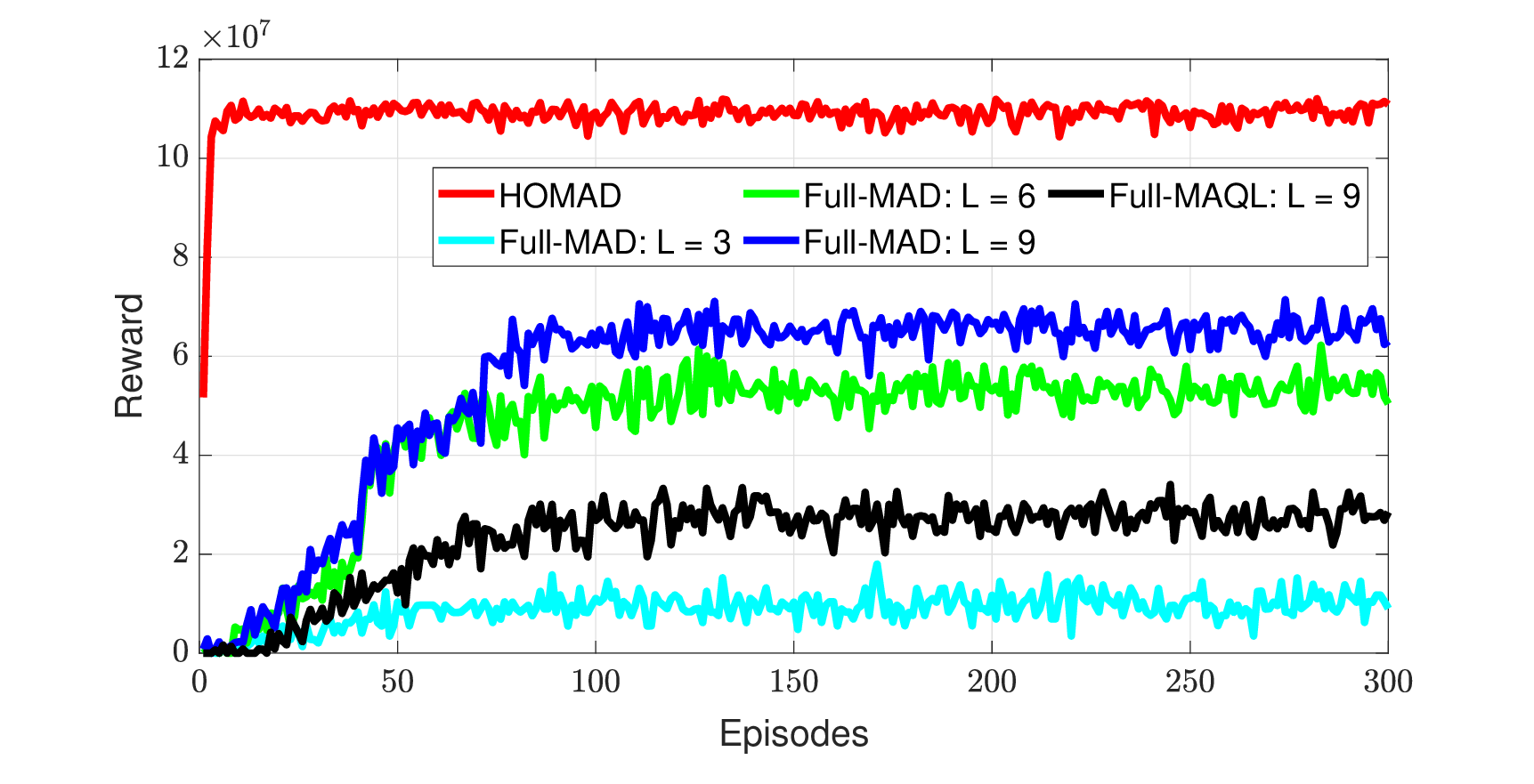}
    \vspace{-10pt}
	\caption{Convergence comparison of different learning methods, where $M_{\sf{U}} = M_{\sf{E}} = 1$, $M_{\sf{M}} = 4$, $K = 2$, $\nu_{\sf{E}} = 1$, $\nu_{\sf{U}} = 4$, $D_{max} = 2$ (ms), $\varepsilon_{th} = 10^{-5}$, and $R_{e}^{tar} = 4$ (bps/Hz).}
	\label{Fig2_conf}
	\vspace{-15pt}
\end{figure}

Fig. \ref{Fig2_conf} depicts the convergence trend during the learning process of full multi-agent Q-learning (Full-MAQL), Full-MAD, and HOMAD approach by illustrating the variation of the reward achieved by all agents versus the various number of episodes. 
In the Full-MAQL scheme, each agent needs to build its own Q-table including all possible sate-action combinations. 
Fig. \ref{Fig2_conf} shows that Full-MAQL returns the lowest reward (i.e., worst performance) in comparison to other schemes. 
This demonstrates the limitation of the Q-learning approach in a very large-space environment. 
Considering our proposed schemes, the HOMAD algorithm can achieve the highest rewards in this simulation with the significant gaps between its corresponding curve and the others thanks to the power-optimization process.
Due to the discrete power level of the quantization process, which is widely used in literature \cite{SilvaWCL20,FayazTWC21}, the Full-MAD scheme obtains a lower reward than that due to the HOMAD. 
Interestingly, the Full-MAD scheme can improve its performance by increasing the number of TPLs ($L$) as shown in Fig. \ref{Fig2_conf} which however leads to a larger action space together with a higher complexity level of the learning process.
The convergence is further clarified in Table \ref{tab:conver_time} where the number of episodes required for convergence, the average implementing time per episode, and the convergence time corresponding to three schemes are provided. As given in this table, 
HOMAD scheme converges with the lowest number of episodes but also requires the highest implementing time per episode because of its smallest action space size and also the power-optimization process. 
Inversely, Full-MAQL needs the largest number of episodes for convergence while its average time for each episode is the shortest. 
In summary, the HOMAD scheme again shows its superiority to the others when requires the shortest time for convergence.  


Next Figs. \ref{Fig3_conf} and \ref{Fig4_conf} illustrate the variation of the average EE versus the different value sets of $\left(D_{max}, \varepsilon_u, R_e^{\sf{tar}} \right)$ and number of mMTC users, respectively. 
In Fig. \ref{Fig3_conf}, we aim to consider a scenario, where mMTC users can use the same SCs assigned to URLLC and eMBB users to improve spectral efficiency and connectivity density, as long as URLLC and eMBB requirements, i.e., $\left(D_{max}, \varepsilon_u, R_e^{\sf{tar}} \right)$, are still guaranteed. As expected, the higher stringent requirements of URLLC and eMBB users (i.e., lower $D_{max}$ and $\varepsilon_u$, and higher $R_e^{\sf{tar}}$) result in the lower EE achieved by all schemes. 
In Fig.~\ref{Fig4_conf}, one can observe that increasing $M_{\sf{M}}$ will degrade the system EE. This is because the number of mMTC users using the same SC gets higher as $M_{\sf{M}}$ increases which results in the higher interference suffering the URLLC and eMBB users. In addition, these figures again confirm the superiority of the proposed HOMAD algorithm in all simulation scenarios, while the Full-MAD algorithm outperforms the Full-MAQL scheme.
\begin{table}[!t]
    \caption{Convergence Time Comparison}
    \vspace{-5pt}
    \centering
    \scriptsize
    \begin{tabular}{|l|c|c|c|}
        \hline
        \textbf{\shortstack{Methods}} & \textbf{\shortstack{Avg. time per episode}} & \textbf{\shortstack{No. of episodes}} & \textbf{\shortstack{Conv. time}}  \\
        \hline
        Full-MAQL & 0.36 sec. & 85 & 31.05 sec. \\
        \hline
        Full-MAD & 1.58 sec. & 85 & 134.46 sec. \\
        \hline
        HOMAD & 2.49 sec. & \textcolor{red}{6} & \textcolor{red}{14.93 sec.} \\
        \hline
    \end{tabular}
    \normalsize
    \label{tab:conver_time}
    \vspace{-10pt}
\end{table}

\begin{figure}[!t]
	\centering
	\includegraphics[width=8cm,height=5cm]{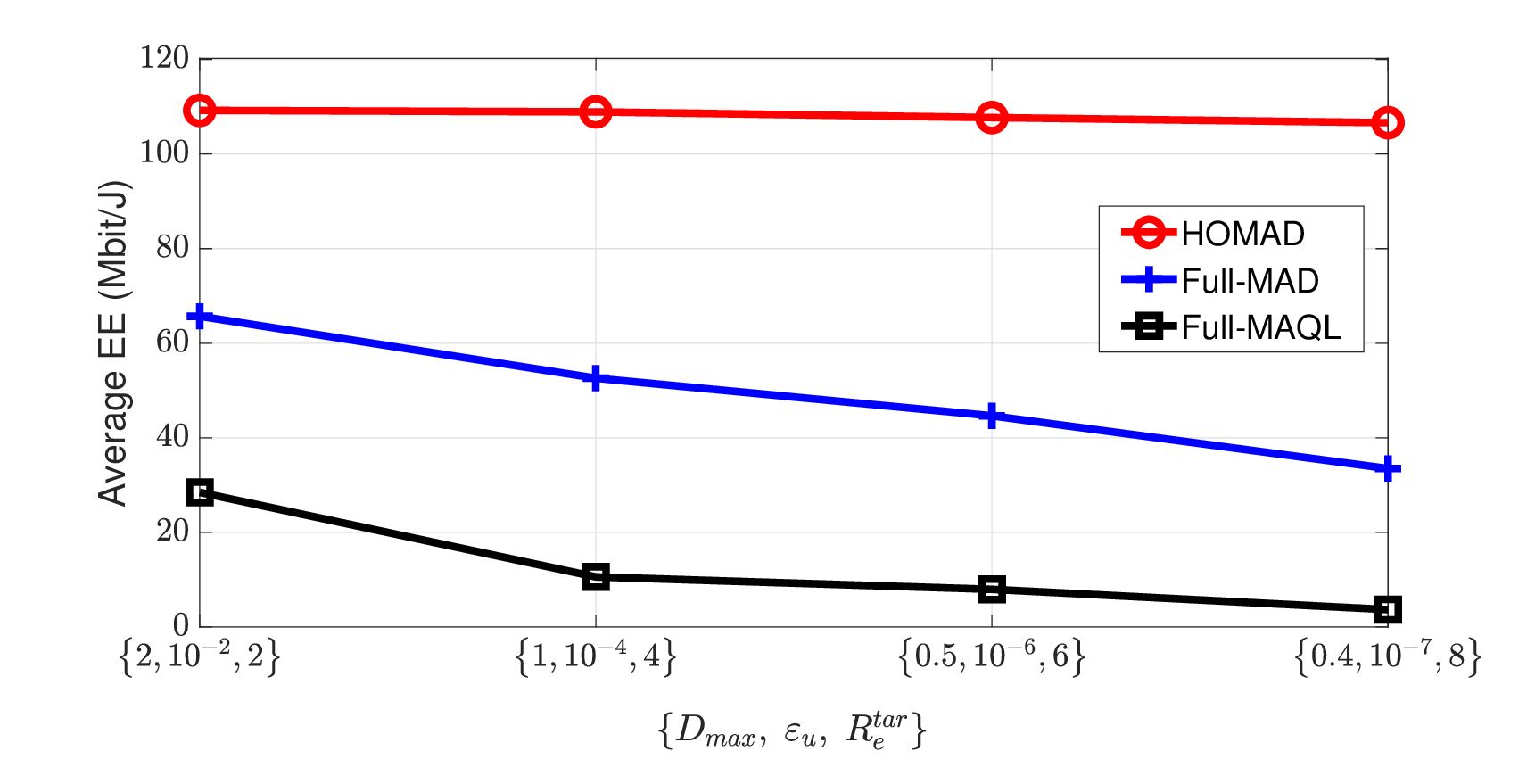}
    \vspace{-10pt}
	\caption{Effect of URLLC and eMBB requirements $\left\{ D_{max}, \; \varepsilon_{u}, \; R_e^{tar} \right\}$, where $M_{\sf{U}} = M_{\sf{E}} = 1$, $M_{\sf{M}} = 4$, $K = 2$, $\nu_{\sf{E}} = 1$, and $\nu_{\sf{U}} = 4$.}
	\label{Fig3_conf}
	\vspace{-15pt}
\end{figure}

\begin{figure}[!t]
	\centering
	\includegraphics[width=8cm,height=5cm]{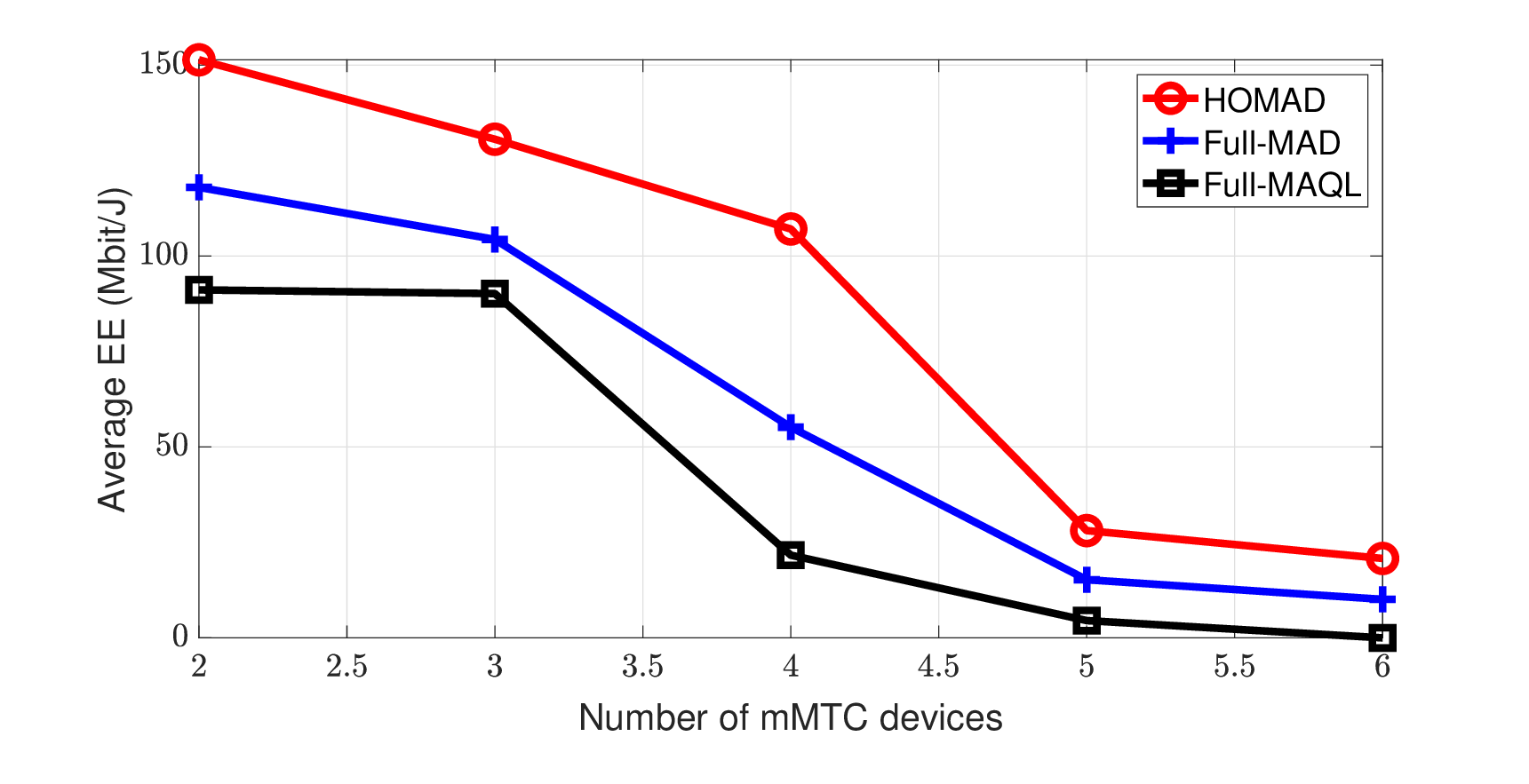}
 \vspace{-10pt}
	\caption{Effect of $M_{\sf{M}}$, where $M_{\sf{U}} = M_{\sf{E}} = 1$, $K = 2$, $\nu_{\sf{E}} = 1$, $\nu_{\sf{U}} = 4$, $D_{max} = 2$ (ms), $\varepsilon_{th} = 10^{-5}$, and $R_{e}^{tar} = 4$ (bps/Hz).}
	\label{Fig4_conf}
	\vspace{-15pt}
\end{figure}

\section{Conclusion}
We have proposed two multi-agent Deep RL-based resource allocation mechanisms, HOMAD and Full-MAD algorithms, for maximizing the system EE of 
eMBB/mMTC/URLLC-coexistence 5G-NR networks using semi-GF NOMA transmission strategy.
In particular, the Full-MAD approach addresses the EE maximization problem by employing the MADQN method to conduct both SC and PA selection. 
Furthermore, the HOMAD approach aims to use the MADQN method to only select SC solution while the power corresponding to a given SC setting can be optimized effectively. 
Simulation results have shown that the Full-MAD method outperforms the conventional Full-MAQL mechanism, while the HOMAD algorithm can return a higher EE and converge faster than other benchmark schemes.

\section*{Acknowledgment}
This work was supported by the National Research Fund (FNR), Luxembourg under the CORE project 5G-Sky (Grant C19/IS/13713801), and ERC-funded project Agnostic (Grant 742648).

\ifCLASSOPTIONcaptionsoff
  \newpage
\fi

\bibliographystyle{IEEEtran}
\bibliography{main}

\begin{thebibliography}{10}
\providecommand{\url}[1]{#1}
\csname url@samestyle\endcsname
\providecommand{\newblock}{\relax}
\providecommand{\bibinfo}[2]{#2}
\providecommand{\BIBentrySTDinterwordspacing}{\spaceskip=0pt\relax}
\providecommand{\BIBentryALTinterwordstretchfactor}{4}
\providecommand{\BIBentryALTinterwordspacing}{\spaceskip=\fontdimen2\font plus
\BIBentryALTinterwordstretchfactor\fontdimen3\font minus
  \fontdimen4\font\relax}
\providecommand{\BIBforeignlanguage}[2]{{%
\expandafter\ifx\csname l@#1\endcsname\relax
\typeout{** WARNING: IEEEtran.bst: No hyphenation pattern has been}%
\typeout{** loaded for the language `#1'. Using the pattern for}%
\typeout{** the default language instead.}%
\else
\language=\csname l@#1\endcsname
\fi
#2}}
\providecommand{\BIBdecl}{\relax}
\BIBdecl

\bibitem{LienCM17}
S.-Y. Lien \emph{et~al.}, ``{5G} new radio: Waveform, frame structure, multiple
  access, and initial access,'' \emph{{IEEE} Commun. Mag.}, vol.~55, no.~6, pp.
  64--71, 2017.

\bibitem{TiTi_ComLet20}
T.~T. Nguyen, V.~N. Ha, and L.~B. Le, ``Wireless scheduling for heterogeneous
  services with mixed numerology in {5G} wireless networks,'' \emph{{IEEE}
  Commun. Lett.}, vol.~24, no.~2, pp. 410--413, 2020.

\bibitem{ShahabCST20}
M.~B. Shahab, R.~Abbas, M.~Shirvanimoghaddam, and S.~J. Johnson, ``Grant-free
  non-orthogonal multiple access for {IoT}: A survey,'' \emph{{IEEE} Commun.
  Surveys Tuts.}, vol.~22, no.~3, pp. 1805--1838, 2020.

\bibitem{DungTran_SJ2020}
D.-D. Tran, H.-V. Tran, D.-B. Ha, and G.~Kaddoum, ``Secure transmit antenna
  selection protocol for {MIMO NOMA} networks over {Nakagami-m} channels,''
  \emph{{IEEE} Syst. J.}, vol.~14, no.~1, pp. 253--264, 2020.

\bibitem{SilvaWCL20}
M.~V. da~Silva, R.~D. Souza, H.~Alves, and T.~Abr{\~a}o, ``A {NOMA}-based
  {Q}-learning random access method for machine type communications,''
  \emph{{IEEE} Wireless Commun. Lett.}, vol.~9, no.~10, pp. 1720--1724, 2020.

\bibitem{DungTranGLOBECOM21}
D.-D. Tran, S.~K. Sharma, S.~Chatzinotas, and I.~Woungang, ``Learning-based
  multiplexing of grant-based and grant-free heterogeneous services with short
  packets,'' in \emph{Proc. {IEEE} Global Commun. Conf.}, 2021.

\bibitem{DungTranPIMRC21}
------, ``Q-learning-based {SCMA} for efficient random access in {mMTC}
  networks with short packets,'' in \emph{Proc. {IEEE} Int. Symp. Pers.,
  Indoor, Mobile Radio Commun.}, 2021, pp. 1334--1338.

\bibitem{DungTran_VTC21}
D.-D. Tran, S.~K. Sharma, and S.~Chatzinotas, ``{BLER}-based adaptive
  {Q}-learning for efficient random access in {NOMA}-based {mMTC} networks,''
  in \emph{Proc. {IEEE} Veh. Technol. Conf.}, 2021, pp. 1--5.

\bibitem{DungTranVTC22}
D.-D. Tran, V.~N. Ha, and S.~Chatzinotas, ``Novel reinforcement learning based
  power control and subchannel selection mechanism for grant-free {NOMA
  URLLC}-enabled systems,'' in \emph{IEEE VTC-Spring}, 2022.

\bibitem{FayazTWC21}
M.~Fayaz, W.~Yi, Y.~Liu, and A.~Nallanathan, ``Transmit power pool design for
  grant-free {NOMA-IoT} networks via deep reinforcement learning,''
  \emph{{IEEE} Trans. Wireless Commun.}, vol.~20, no.~11, 2021.

\bibitem{FayazArxiv22}
------, ``A power-pool-based power control in semi-grant-free {NOMA}
  transmission,'' \emph{arXiv preprint arXiv:2106.11190v2}, pp. 1--14, 2022.

\bibitem{LiuTCOM23}
Y.~Liu, Y.~Deng, H.~Zhou, M.~Elkashlan, and A.~Nallanathan, ``Deep
  reinforcement learning-based grant-free {NOMA} optimization for {mURLLC},''
  \emph{{IEEE} Trans. Commun.}, vol.~71, no.~3, pp. 1475--1490, 2023.

\bibitem{DungTranOJCOM23}
D.-D. Tran, S.~K. Sharma, V.~N. Ha, S.~Chatzinotas, and I.~Woungang,
  ``Multi-agent {DRL} approach for energy-efficient resource allocation in
  {URLLC}-enabled grant-free {NOMA} systems,'' \emph{{IEEE} Open J. Commun.
  Soc.}, pp. 1--17, 2023, early access.

\bibitem{VuHaNL20}
V.~N. Ha, T.~T. Nguyen, L.~B. Le, and J.-F. Frigon, ``Admission control and
  network slicing for multi-numerology {5G} wireless networks,'' \emph{{IEEE}
  Netw. Lett.}, vol.~2, no.~1, pp. 5--9, 2020.

\bibitem{HanTVT19}
S.~Han \emph{et~al.}, ``Energy-efficient short packet communications for uplink
  {NOMA}-based massive {MTC} networks,'' \emph{{IEEE} Trans. Veh. Technol.},
  vol.~68, no.~12, pp. 12\,066--12\,078, December 2019.

\bibitem{DingCOML20}
Z.~Ding, R.~Schober, and H.~V. Poor, ``Unveiling the importance of {SIC} in
  {NOMA} systems-part 1: State of the art and recent findings,'' \emph{{IEEE}
  Commun. Lett.}, vol.~24, no.~11, pp. 2373--2377, 2020.

\bibitem{DungTran_TVT2021}
D.-D. Tran, S.~K. Sharma, S.~Chatzinotas, I.~Woungang, and B.~Ottersten,
  ``Short-packet communications for {MIMO NOMA} systems over {Nakagami-m}
  fading: {BLER} and minimum blocklength analysis,'' \emph{{IEEE} Trans. Veh.
  Technol.}, vol.~70, no.~4, pp. 3583--3598, 2021.

\bibitem{Sun2019TWC}
C.~Sun \emph{et~al.}, ``Optimizing resource allocation in the short blocklength
  regime for ultra-reliable and low-latency communications,'' \emph{{IEEE}
  Trans. Wireless Commun.}, vol.~18, no.~1, pp. 402--415, Jan. 2019.

\bibitem{VuHaTGCN20}
V.~N. Ha, D.~H.~N. Nguyen, and J.-F. Frigon, ``System energy-efficient hybrid
  beamforming for mmwave multi-user systems,'' \emph{{IEEE} Trans. Green
  Commun. Netw.}, vol.~4, no.~4, pp. 1010--1023, 2020.

\bibitem{VuHa_ICC18}
------, ``Energy-efficient hybrid precoding for mmwave multi-user systems,'' in
  \emph{Proc. {IEEE} Int. Conf. Commun.}, 2018, pp. 1--6.

\bibitem{VuHa_TWC21}
V.~N. Ha, G.~Kaddoum, and G.~Poitau, ``Joint radio resource management and link
  adaptation for multicasting 802.11ax-based {WLAN} systems,'' \emph{{IEEE}
  Trans. Wireless Commun.}, vol.~20, no.~9, pp. 6122--6138, 2021.

\bibitem{BoydBook}
S.~Boyd and L.~Vandenberghe, \emph{Convex Optimization}.\hskip 1em plus 0.5em
  minus 0.4em\relax Cambridge University Press, 2004.

\end{thebibliography}

\end{document}